\documentclass[12pt]{article}

\usepackage{graphicx}
\usepackage{times}
\usepackage{tikz}
\usepackage{cite}

\usepackage{siunitx}  % Added by JA: useful for getting units upright

\usepackage[normalem]{ulem}% Added by Francis for the \sout command

\newenvironment{sciabstract}{%
\begin{quote} \bf}
{\end{quote}}

\title{Nonlinear contractile response of actomyosin active gels to control signals}% Force line breaks with \\
\author
{James Clarke$^{1\,\dagger}$, Francis Cavanna$^{1\,\dagger}$, Aniket Marne$^{1}$,\\ Anthony Davolio$^{1}$, Jos\'{e} Alvarado$^{1\ast}$\\
\\
\normalsize{$^{1}$Center for Nonlinear Dynamics, Department of Physics,} \\[0.5ex] \normalsize{The University of Texas at Austin. Austin, TX, USA}\\
\\
\normalsize{$\dagger$ : Authors contributed equally.}
\\
\normalsize{$^\ast$ : To whom correspondence should be addressed; E-mail:  alv@chaos.utexas.edu.}
}

\date{}

\begin{document} 


\baselineskip24pt


\maketitle


\begin{sciabstract}
Biological systems tightly regulate their physiological state using control signals. This includes the actomyosin cytoskeleton, a contractile active gel that consumes chemical free energy to drive many examples of cellular mechanical behavior. Upstream regulatory pathways activate or inhibit actomyosin activity. However, the contractile response of the actomyosin cytoskeleton to control signals remains poorly characterized. Here we employ reconstituted actomyosin active gels and subject them to step and pulsatile activation inputs. We find evidence for a nonlinear impulse response, which we quantify via a transfer function $\delta \varepsilon / \delta g$ that relates input free-energy pulses $\delta g$ to output strain pulses $\delta \varepsilon$. We find a scaling relation $\delta \varepsilon / \delta g \sim g^{-0.3}$. The negative sign of the exponent represents a decreased effectiveness of a contracting gel in converting energy to strain. We ascribe nonlinearity in our system to a density-dependent mechanism, which contrasts strain-stiffening nonlinear responses to external stresses. Contractile response to control signals is an essential step toward understanding how information from mechanical signaling processes flow through actomyosin networks in living, and likely also synthetic, cells.

\end{sciabstract}

\section*{Introduction}

Cells are micron-scale acrobats, capable of remarkable mechanical tasks such as crawling, dividing, and healing wounds. To describe these tasks, simple mechanical models often suffice.\cite{isomursuDirectedCellMigration2022} But mechanics alone are not enough.  Cells also rely on intracellular biochemical signaling proteins, which constitute an information-processing “circuit” that drives gene expression and thus decision-making within and across cell cycles.\cite{dueberRewiringCellSignaling2004} A well-studied example of signaling in a mechanical context is mesenchymal stem-cells, whose fate is sensitive to mechanical information.\cite{englerMatrixElasticityDirects2006,debellyInterplayMechanicsSignalling2022} A tight orchestration between mechanics and signaling is required to coordinate proper cellular acrobatics and, more broadly, survival.

One key driver of cellular mechanical behavior is the actomyosin cytoskeleton. Its mechanical properties and ability to contract have been well studied.\cite{murrellForcingCellsShape2015} Equally important is its interactions with biochemical signaling pathways. Actomyosin activity responds to upstream signals that are mediated by Rac1 and RhoA \cite{tsujiROCKMDia1Antagonize2002} and calcium.\cite{somlyoCa2SensitivitySmooth2003} Meanwhile, actomyosin forces contribute to downstream signals triggered by mechanosensitive proteins such as filamin,\cite{huOpposingFlnAFlnB2017} integrin complexes,\cite{kechagiaIntegrinsBiomechanicalSensors2019} and ion channels.\cite{ellefsenMyosinIIMediatedTraction2019} Furthermore, actomyosin tension affects the binding affinity of accessory proteins, potentially also contributing to intracellular signaling.\cite{sunMechanosensingDirectBinding2020} Finally, mechanical information influences several key cellular pathways, including Wnt,\cite{steinhartWntSignalingDevelopment2018} YAP/TAZ,\cite{totaroYAPTAZLink2017} and Hippo.\cite{yuHippoPathwayOrgan2015} In all, actomyosin contractility is deeply embedded in the intracellular signaling circuitry. But how does contractility participate in information flows within this circuit?

Control theory and information theory have offered powerful quantitative frameworks for describing intracellular information flows. A well researched example is sensing chemical environments. Examples include chemotaxis,\cite{mattinglyEscherichiaColiChemotaxis2021} quiescence/growth,\cite{kramerMultimodalPerceptionLinks2022} and quorum sensing.\cite{mehtaInformationProcessingSignal2009} In these examples, the response (gene expression) to signals (environmental ligand binding) is well understood. Meanwhile for the actomyosin cytoskeleton, the contractile response of actomyosin contractility to activation signals remains less understood. Characterizing this response is necessary to understand how information flows through the actomyosin cytoskeleton in the same way it flows through signaling proteins subject to reaction-diffusion dynamics.

Here we perform experiments with reconstituted gels to determine the contractile response of actomyosin active gels to control signals. We first investigate gels  under an effectively constant activation signal (step response) and observe fast dynamics to a densely contracted configuration. Remarkably, the gel consumes chemical free energy well after contraction ceases. We develop a linear hydrodynamic model, and find disagreement with our observations. This comparison suggests that the gel’s response to motor activity depends on density, leading to nonlinear contractile properties. In order to determine contractile properties, we next investigate pulsatile activation signals (impulse response). Pulsatile contractility has been widely reported across cells,\cite{coravosActomyosinPulsingTissue2017a} yet many experimental studies on actomyosin contractility investigate step responses. Furthermore, an advantage of impulse responses is that they directly probe the system dynamics.\cite{bechhoeferControlTheoryPhysicists2021b} We report a nonlinear contractile energetic compliance $\delta \varepsilon(g)/\delta g$, with a negative scaling with the free-energy density $g$ consumed by the gel. The negative sign of this exponent indicates that as the gel deforms in response to an input of energy $\delta g$, the deformation to the next energy input $\delta g$ will decrease. Nonlinear contractility may lead to concrete advantages. We discuss how pulsatile activity maximizes the energetic cost of contraction. We also discuss the actomyosin as a conduit of signals and mutual information.

\section*{Results}

\subsection*{Step response}

\begin{figure}
    \centering
    \includegraphics[width=\linewidth]{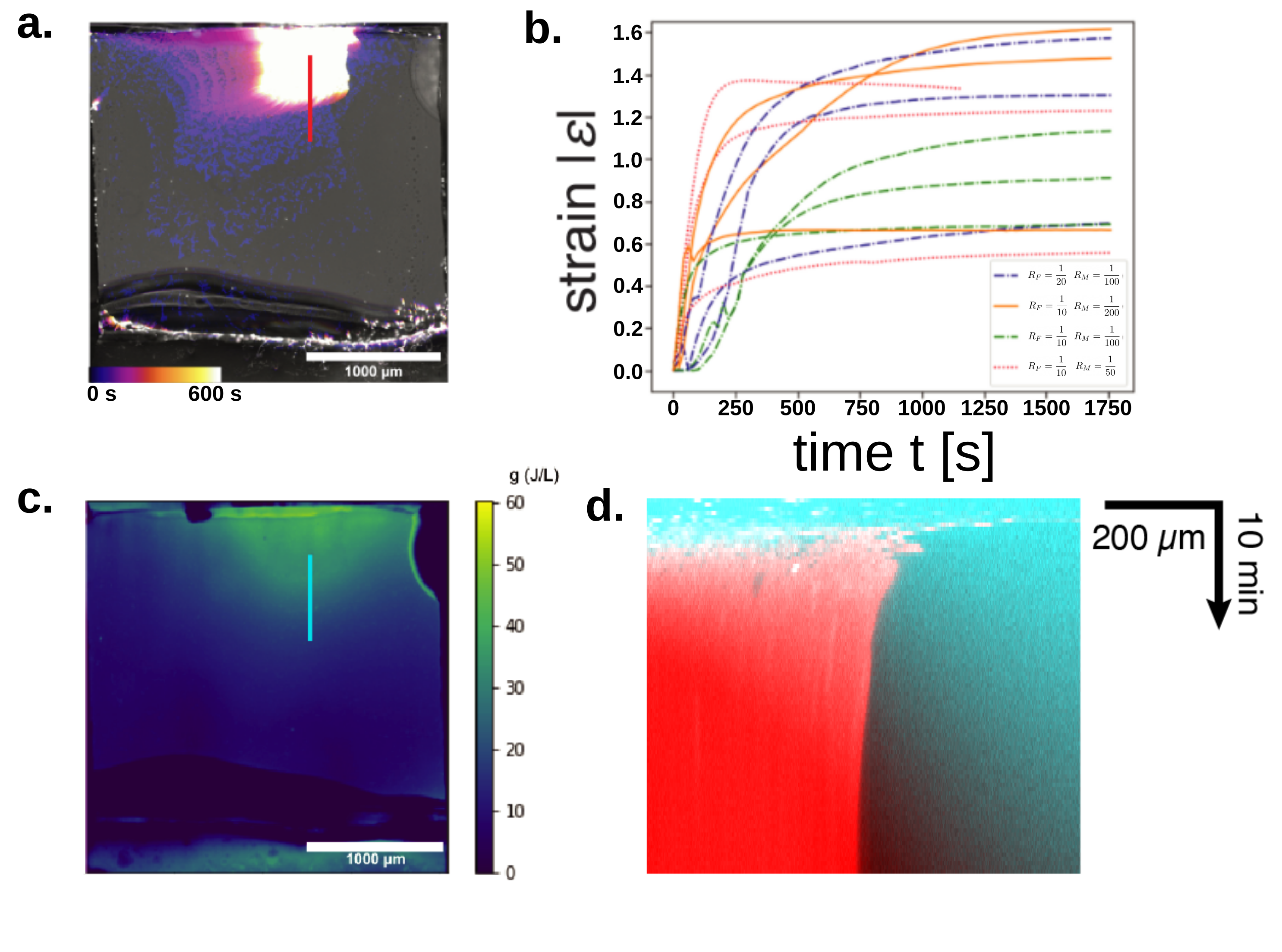}
\caption{Step input activation of actomyosin contraction. a) Color-time overlay of a contraction event ($[actin] = \qty{12}{\micro M}$, $R_M = 1/50$, $R_F = 1/10$). Color corresponds to time after sample preparation (calibration bar, lower left). b) Strain trajectories $\varepsilon(t)$ for twelve independently prepared samples. c) Spatial distribution of cumulative Gibbs free energy consumed $g$, determined by NADH assay. d) Kymograph taken over a column of pixels from the actin (red) and NADH (cyan) images (red line, panel a; cyan line, panel c). The NADH kymograph exhibits a wave of NADH de-fluorescence emanating from the contracted gel.}
    \label{fig:step}
\end{figure}

In order to quantify the contractile response of actomyosin active gels to control signals, we turn to reconstituted active gels of actin filaments, myosin motors, and fascin crosslinks. We first consider a common scenario: contraction in the presence of a constant supply of ATP. This scenario mimics the intracellular environment and is straightforward to implement in reconstituted systems by including enzymes that regenerate ATP.\cite{chenEnzymaticRegenerationConservation2021} We prepare contractile active gels, load onto observation chambers, and record contraction events using fluorescence microscopy (Methods; Fig.(\ref{fig:step}a)). We determine each sample’s strain trajectory $\varepsilon (t)$ by tracking the gels’ imaged area and computing strain (Methods; Fig.(\ref{fig:step}b)). We observe that strain rises from an initial value of zero to some finite, steady value, with $\varepsilon_{f} = 1.10 \pm 0.32$ across all trials. Strain rises within a characteristic timescale $\tau = 220 \pm 130\,\unit{s}$. The strain trajectory appears to resemble an exponential rise $\varepsilon(t) = 1-e(-t/\tau)$, which is the solution to a first-order linear differential equation. This resemblance appears to suggest that contractility agrees with a linear model. 

In order to test this suggestion, we apply a linear model $$0 = K\varepsilon + \eta\dot\varepsilon + J\,,$$ where $K$ is the bulk modulus, $\eta$ is the effective viscosity, and $J$ is a constant active stress exerted by the ensemble of myosin motors. The solution $\varepsilon(t)$ of this equation is given by an exponential rise, with $\varepsilon_{f} = J/K$ and timescale $\tau = \eta/K$. We plug in values based on prior studies to determine agreement between the linear model and experiments. With $K \approx \qty{1}{\pascal}$ \cite{alvaradoMolecularMotorsRobustly2013} assuming that bulk and shear moduli are comparable), $\eta \approx \qty{1}{\pascal\second}$,\cite{lielegSlowDynamicsInternal2011} and $J \approx \qty{30}{\pascal}$,\cite{bendixQuantitativeAnalysisContractility2008a} we have $\varepsilon_{f}\approx30$ and $\tau \approx \qty{1}{\second}$. Comparing these values to our experimental results reveals a discrepancy of two orders of magnitude: our gels deform less and deform more slowly than predicted by the linear model. What is the origin of this discrepancy?

Characterizing the response of active gels to motor activity is experimentally challenging. Measuring the response to external shear is straightforward, because both shear stress $\sigma$ and shear strain $\gamma$ can be simultaneously measured with a rheometer to yield the shear modulus $G = \sigma/\gamma$. Meanwhile, in active gels, measuring the active stress $J$ is more challenging than measuring the bulk strain $\varepsilon$. A few studies have experimentally estimated active stresses for actomyosin,\cite{bendixQuantitativeAnalysisContractility2008a} and microtubule-kinesin active gels.\cite{adkinsDynamicsActiveLiquid2022a} Rather than estimate active stress, here we leverage an established assay to measure consumption of chemical free energy by the myosin motors.\cite{cruzKineticMechanismRegulation2001,chenATPConsumptionEukaryotic2015,sakamotoFactinArchitectureDetermines2024} In short, we couple ATP consumption to the oxidation of fluorescent nicotinamide adenine dinucleotide (NADH) using two enzyme-substrate pairs. Since ATP hydrolysis is stoichiometrically coupled to NADH oxidation, loss of NADH fluorescence reports free-energy consumption. This method allows us to measure the spatial distribution of free-energy consumption while measuring strain simultaneously (Fig.(\ref{fig:step}c)).

A simple expectation is that the active stress $J$ is proportional to the total free-energy density consumed: $J \sim g$. If this assumption were correct, we would expect a constant, finite amount of free-energy consumed by the gel, and this consumption could likely occur over the timescale $\tau = \qty{250}{s}$ of contraction. To test this expectation, we plot a kymograph of the NADH concentration near the site of the contracted cluster (Fig.(\ref{fig:step}d)). We observe that NADH continues to be oxidized well past the characteristic contraction timescale $\tau$, in contrast with our simple expectation. We instead observe a wave of oxidation emanating from the contracted gel. This wave is likely governed by reaction-diffusion dynamics. To confirm, we simulate our experiments by numerically solving for the governing reaction-diffusion dynamics, incorporating a Michaelis-Menten model of reaction kinetics, diffusion of reagents, deformation of the gel, and the assumption that the ATPase rate $k_{myo}$ of the gel is constant. We fit the outputs of multiple reaction-diffusion simulations, allowing $k_{myo}$ and the effective diffusion constants to vary, compare the resulting NADH concentration fields to our data, and minimize the error using the Nelder-Mead method to extract the myosin ATPase rate. Overall, we find good agreement between data and simulation (Fig.(\ref{fig:step}a,b)), with $k_{myo} = 10^{-4}\,\unit{mM\per\second}$. This fit confirms that $\dot g \approx const$, with a typical scale of $\dot g \approx \qty{5}{W/m^3}$. These results are largely consistent with recent measurements.\cite{sakamotoFactinArchitectureDetermines2024}

To summarize, we have experimentally found that (a) strain rises toward a terminal value within a finite time, (b) free energy is continuously consumed well past this time, and (c) a linear model fails to predict values for $\varepsilon_{f}$ and $\tau$. Furthermore, the observation $\dot g = const$ appears to be at odds with $J \sim g$. If both relations were true, active stress would continue to increase with time. Here we consider one modification to our linear model that may reconcile both observations: a density-dependent nonlinearity. As the gel contracts, the density of actin as well as myosin motors increases. We presume this increased density affects $K$, $\eta$, and $J$, thus making them functions of the output strain $\varepsilon$. (We note that in the actomyosin cortex of living cells, actin filaments undergo rapid turnover; nonlinearity in $K$ could thus be excluded \emph{in vivo}; see Discussion.) Indeed, density-dependent mechanisms have been discussed in active gel models \cite{kruseGenericTheoryActive2005,banerjeeInstabilitiesOscillationsIsotropic2011,banerjeeActomyosinPulsationFlows2017b,idesesSpontaneousBucklingContractile2018} but remain poorly characterized experimentally. It may be likely that our step-response data could probe density-dependent nonlinearity. Indeed, an exponential fit to our strain trajectories reveals systematic residues, indicating the presence of additional effects not captured by a linear model. However, measuring density-dependent nonlinearity with a step response is difficult, because deviations from linearity manifest primarily in transients. Probing nonlinear responses in externally driven materials is possible with equipment such as rheometers, which can drive the system sinusoidally with a single frequency and systematically vary that frequency. Active gels cannot be driven “backwards” because active stresses primarily cause contraction, and not extension. Thus, in order to systematically probe the nonlinear contractile response of actomyosin, we turn to a different kind of driving.

\begin{figure}
    \centering
    \includegraphics[width=0.9\linewidth]{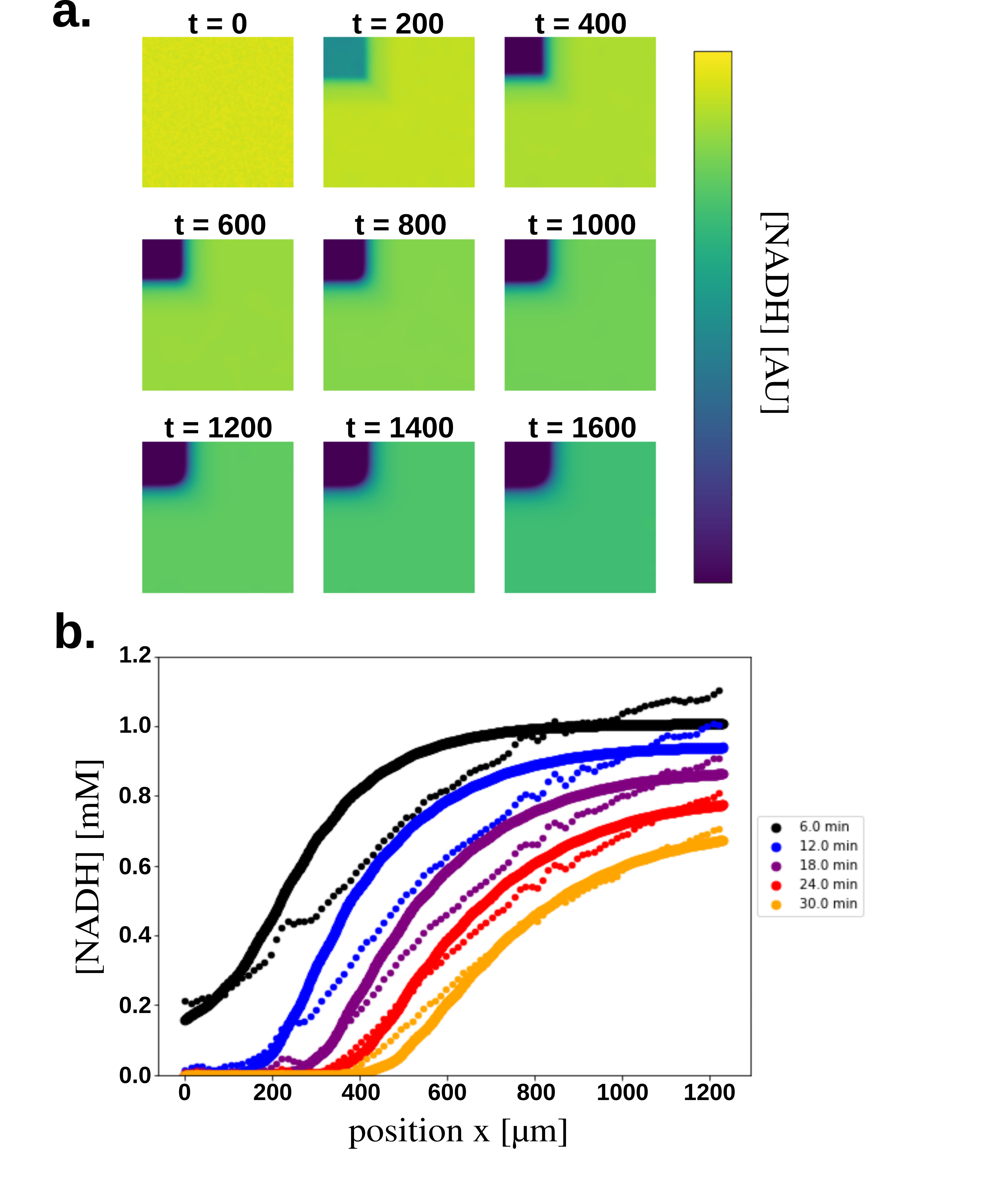}
\caption{Reaction-diffusion dynamics confirm a constant chemical free-energy consumption rate. a) Snapshots of the NADH concentration field for nine sample times: $t=\qty{0}{s}$ (top left) through $t=\qty{1600}{s}$ (bottom right), in steps of $\qty{200}{s}$. Color corresponds to NADH concentration (calibration bar, right). b) Plot of NADH concentration versus space (abscissa) and time (color; legend, right) for experiment (points) and simulation (bold line).}
    \label{fig:step_sim}
\end{figure}

\subsection*{Impulse response}

In biology, actomyosin contractility is rarely activated for sustained periods of time. Researchers have long documented the observation that contractility proceeds in small, spatially and temporally localized “bursts” and “waves”, collectively termed “pulsatility”.\cite{martinPulsedContractionsActin2009,coravosActomyosinPulsingTissue2017a,bairdLocalPulsatileContractions2017} Despite this biological relevance, it remains poorly understood how active gels contract in response to pulses in activity.

\begin{figure}
    \centering
    \includegraphics[width=0.75\textwidth]{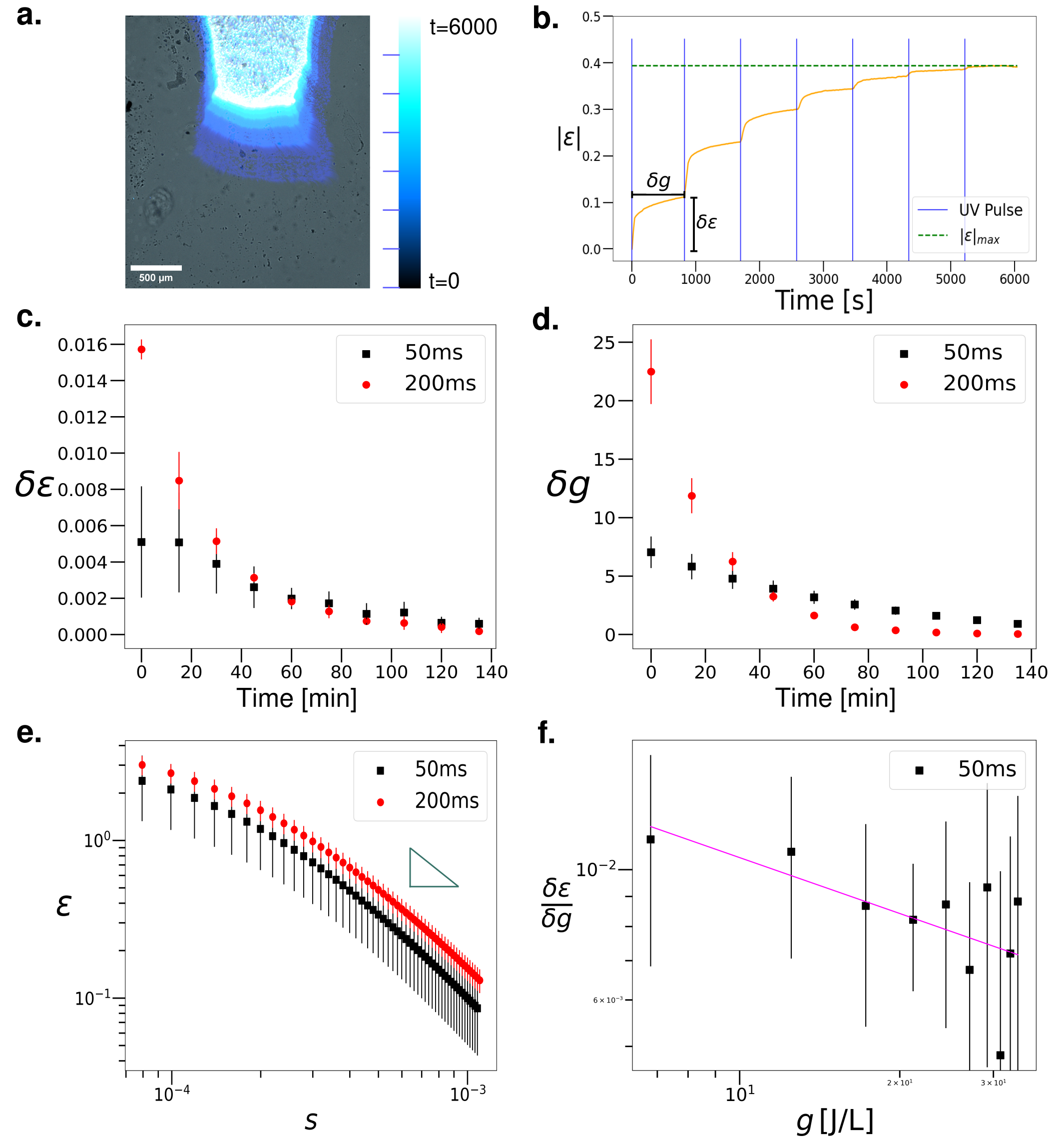}
    \caption{(a) Color-time overlay of epifluorescence micrographs. Color corresponds to time as shown in the color bar to the right. Scale bar = $\qty{500}{\micro\meter}$. (b) Absolute value of strain vs. time for a representative sample. Blue vertical bars in overlay represent UV pulse inputs that correspond to the blue horizontal lines alongside the color bar in panel a. Maximum strain achieved in this sample is indicated by the horizontal dashed green line. (c) The change in strain, $\delta\varepsilon$, as a function of time, $t$, shown for $\tau_e = \qty{50}{\milli\second}$ (black squares), and $\tau_e = \qty{200}{\milli\second}$ (red circles).(e) The change in the amount of consumed free energy, $\delta g$ as a function of time, $t$, shown in the same format as panel c. (e) Laplace transformed strain, $\varepsilon$, vs. Laplace frequency $s$, shown in the same format as panel c. Inset right triangle indicate the high frequency scaling with $\xi_{50}=-1.783\pm0.002$, and $\xi_{200}=-1.678\pm0.002$. (f) $\delta\varepsilon/\delta\,g$ vs. cumulative free energy density $g$ shown on loglog axes. Magenta line indicates the linear fit of the data with a slope of $\zeta = -0.316\pm0.128$.}
    \label{fig:impulse}
\end{figure}

We drive actomyosin gels impulsively using an assay including caged ATP (Methods). This analog of ATP remains protected from hydrolysis until illuminated with UV (365 nm) light, which induces an “uncaging” reaction, thereby making the ATP available for hydrolysis. We prepare the biochemical composition of these gels identically (Methods); we instead vary across experiments the pulse width, $\tau_e$, between 50 and 200 $\unit{ms}$. We chose this range because we found that $\qty{50}{ms}$ pulses are short enough to liberate a small fraction of caged ATP, while $\qty{200}{ms}$ pulses uncaged a majority of all caged ATP molecules. We wait $\qty{900}{s}$ between pulses, which is significantly longer than the rise time of $\tau^* = \qty{250}{s}$ we observed in the step assay. We again use fluorescence microscopy to observe the contraction resulting from each pulse. Fig.(\ref{fig:impulse}a,b) demonstrate the discrete contractile events that result from individual activation pulses - where panel a provides a time-color overlay, and panel b plots the strain evolution as a function of time, $\varepsilon(t)$, as shown by the orange curve. The individual pulses (vertical blue lines) differ from the step response: rather than immediately rising to a single terminal strain $\varepsilon_f$, each pulse releases a small packet $\delta g$ of energy that ultimately results in a small increase $\delta \varepsilon$ of strain (Fig.(\ref{fig:impulse}c)).

To determine $\delta g$, we develop an analytical model accounting for Michaelis-Menten kinetics (Methods). The results of this model are represented in Fig.(\ref{fig:impulse}d) where it is in good agreement with the $\delta\varepsilon$ data shown in Fig.(\ref{fig:impulse}c). We keep the pulse duration constant within an experiment. However, each pulse releases ever fewer caged ATP molecules and thus $\delta g$ per pulse decreases. This is because this assay lacks ATP regeneration as in the step assay. After several pulses, gels in the pulse assay approach an ultimate terminal strain $\varepsilon_f$ (Fig.(\ref{fig:impulse}b)), horizontal green line). We note that the values of terminal strain observed in our pulse assay are lower than those in the step response. We carefully choose our starting concentration of caged ATP to be sufficiently low to exclude steric repulsion from close-packing of actin, which likely plays a significant role once the gel is fully contracted. 

% This is evident in Figure 3d, where we observe that the normalized strain evolves approximately linearly with the normalized amount of consumed free energy, consistent with the traction regime. 
% Further evident in Figure 3d is that we observe a similar collapse as before for our pulses in the normalized strain and energy consumption $g$ when rescaling the different experiment series with varying pulse times. This indicates that the underlying dynamics of the system are captured in this relationship, as in the case of step input.

Although there appears to be a linear relationship between strain and energy consumption, we ask whether the dynamics are indeed governed by a linear contractile compliance. In order to answer this question, we take inspiration from a quantity developed in nonlinear rheology: the differential modulus $\delta \sigma / \delta \gamma$.\cite{gardelElasticBehaviorCrossLinked2004} In analogy, we ask whether the differential contractile compliance, $\delta \varepsilon / \delta g$, is linear or nonlinear.

To answer this question, we first turn to the frequency dependence of $\varepsilon$ in Laplace space (Fig.(\ref{fig:impulse}e)). We observe a functional form consistent with that of a low-pass filter, expected of many dynamical systems.\cite{bechhoeferControlTheoryPhysicists2021b} We fit to the functional form $\varepsilon = \frac{\varepsilon_f}{1 + s^\xi}$, where $s$ is a Laplace frequency scaled by the filter’s corner frequency, $\xi$ is a scaling exponent, and $\varepsilon_f$ is the terminal strain, or zero-frequency response known as the DC gain. We obtain values for $\xi$ of $\xi_{50}=-1.783\pm0.002$, and $\xi_{200}=-1.678\pm0.002$. These values of $\xi$ indicate that the strain behavior of the system is not a linear first-order system, as expected for the case when $\xi\neq1$.

% If the dynamics of $C$ were linear and purely first-order, we would find $\xi = 1$. Interestingly, we find that this is the case for the shortest pulse window duration of $\qty{50}{ms}$. However, pulses $\qty{100}{ms}$, $\qty{200}{ms}$, and $\qty{500}{ms}$ pulse durations yield a larger value of alpha (Figure callout). This result shows that delivering energy quasistatically can yield linear, first-order dynamics in $C$ if the energy packet is small enough. However, larger packets of energy result in nonlinear dynamics, as reflected in the larger value of $\xi$.

To further test for linearity, we investigate the dependence of the compliance, $\delta\varepsilon/\delta g$, on $g$. If this relationship were linear, it would have no dependence on $g$. However, we indeed find a scaling relationship $\delta\varepsilon/\delta g \sim g^\zeta$, with $\zeta = - 0.316\pm0.128$ (Fig.(\ref{fig:impulse}f)). We note that this nonlinear scaling occurs at all values of $g$ that we investigated. This is in sharp contrast to previously reported mechanical nonlinear quantities, which exhibit a linear regime at low driving and a nonlinear regime above a threshold stress or strain onset,\cite{stormNonlinearElasticityBiological2005} or critical point.\cite{sharmaStraincontrolledCriticalityGoverns2016} The lack of such an onset here agrees with the idea of a density-dependent nonlinear contractile response. The negative sign of the scaling exponent $\zeta$ reflects diminishing returns: as the gel contracts in response to a packet $\delta g$ of energy, it contracts less when receiving an additional packet $\delta g$.

\section*{Discussion}

\subsection*{Potential mechanisms underlying nonlinearity}

We have quantified the contractile response of contractile active gels to two kinds of input signals: step signals and pulsatile signals. We found that the impulse response, characterized by the quasistatic transfer function $\delta \varepsilon/\delta g$, varied with the total amount of free energy density $g$ consumed by the gel (cf. Fig.(\ref{fig:impulse}f)). This nonlinear response demonstrates that the contractile properties of actomyosin contractile gels change as they contract. This nonlinearity is reminiscent of nonlinear responses to external shear in biopolymer networks.\cite{gardelElasticBehaviorCrossLinked2004,stormNonlinearElasticityBiological2005,sharmaStraincontrolledCriticalityGoverns2016} The main difference is that we quantify responses to energy consumption (and, by inference, to active stress) from internal myosin ATPase activity, rather than external stresses. This response is more closely aligned with the actomyosin cytoskeleton’s primary function of generating tension and shape change. Another significant difference is that the nonlinearity underlying contractile response is density-dependent. Biopolymer networks typically strain-stiffen above a critical strain.\cite{gardelElasticBehaviorCrossLinked2004,stormNonlinearElasticityBiological2005} Meanwhile, density-dependent nonlinearity emerges even at small inputs. Density-dependent nonlinear properties were considered in prior theoretical studies \cite{banerjeeInstabilitiesOscillationsIsotropic2011,banerjeeActomyosinPulsationFlows2017b} and predicted contractile regimes (“contraction instability”).

We propose a simple picture to explain the nonlinear response of contractile actomyosin gels. We regard the actin filaments of the gel comprising two populations: the \textit{tension-bearing backbone} and the \textit{residual background network}. During the initial stages of contraction, many actin filaments undergo tension from myosin motor activity. Prior studies have shown that motors bend, buckle, fragment, and compact actin.\cite{murrellFactinBucklingCoordinates2012,vogelMyosinMotorsFragment2013} Therefore, we anticipate that as contraction proceeds, actin filaments gradually transition from a tension-bearing state (before being processed by myosin) to a space-filling medium that no longer bears tension. When contraction ceases, the terminal strain is likely determined by a balance between active stress $J$ of the remaining tension-bearing backbone and the bulk compressibility $K$ of the residual background network. A recent study has demonstrated that interfilament contacts contribute significantly to external compression in actin systems, similar to sheep’s wool;\cite{bouzidTransientContactsFilaments2024} we thus expect that $K$ is set by such interfilament contacts. Furthermore, $J$ is likely subject to additional effects beyond the motor distance, including transverse interfilament tension propagation.\cite{lenzGeometricalOriginsContractility2014} It would be interesting to study how these two concurrent mechanisms give rise to nonlinear response. However, directly measuring active stress remains an experimental challenge, and accurately extracting thermodynamic quantities like mechanical work and power remain a challenge.

Additional physical mechanisms may also contribute to density-dependent nonlinear contractile response. Crosslinks of different types affect the architecture of the actin network,\cite{koenderinkArchitectureShapesContractility2018} which in turn affects the response to external stress,\cite{lielegStructureDynamicsCrosslinked2010} and myosin energy conversion to work.\cite{sakamotoFactinArchitectureDetermines2024} Additionally, our model does not account for dissipation due to elastic stress relaxation,\cite{floydQuantifyingDissipationActomyosin2019} or plasticity.\cite{molnarPlasticCellMechanical2021} Additionally, our study assumes that the coupling between free energy $g$ and active stress $J$ is constant. Although this is a common assumption, its validity is difficult to test experimentally. Molecular motors deliver limited power, which is captured in the force-velocity curve. Furthermore, assemblies of motors, such as muscle and myoblasts, can exhibit a nonlinear force-velocity curve,\cite{hillHeatShorteningDynamic1938,mitrossilisSinglecellResponseStiffness2009} which can result from a nonlinear coupling between myosin unbinding and force/velocity.\cite{piazzesiSkeletalMusclePerformance2007b,seowHillEquationMuscle2013b}

\subsection*{Extending to cells, tissues, and robots}

One difference between our \textit{in-vitro} assay and the actomyosin cytoskeleton of living systems lies in dynamics and adaptability. The actomyosin cytoskeleton is highly dynamic and undergoes constant assembly, restructuring, and disassembly dependent on the mechanical task. Actin filaments \textit{in-vivo} undergo turnover over timescales of a few minutes,\cite{fritzscheAnalysisTurnoverDynamics2013} allowing for faster stress relaxation than our in-vitro assay. We expect that any residual background network \textit{in-vivo} would undergo rapid turnover. This would result in a negligible bulk modulus $K$. We would therefore expect a scaling relation $\delta \varepsilon/ \delta g \sim g^\zeta$ with some positive scaling exponent $\zeta$ in-vivo. The positive sign would reflect the fact that active stress $J$ increases as contraction increases motor density.

However, for timescales shorter than actomyosin turnover, we may still expect a negative sign for $\zeta$. We have shown that the transfer function $\delta\varepsilon/\delta g$ is maximal in the limit $g \rightarrow 0$ due to the negative sign of the exponent $\zeta$ in the scaling relation $\delta \varepsilon / \delta g \sim g^\zeta$ (cf. Fig.(\ref{fig:impulse}f)). This is because an influx of energy $\delta g$ causes an increase in density, which affects the gel in a way that makes future supplies of energy $\delta g$ less effective: the negative sign of $\zeta$ therefore represents diminishing returns. The transfer function $\delta \varepsilon / \delta g$ has an alternative interpretation: it quantifies the energetic cost of contraction, akin to economy measures like cost of transport. \cite{alexanderModelsScalingEnergy2005} In this context, our results suggest that energy economy is maximized in cells when the cytoskeleton is activated with pulsed signals, as opposed to step signals (provided that pulses occur less frequently than the timescale of turnover of the actomyosin cytoskeleton).

Understanding the actomyosin cytoskeleton’s response to control signals is an essential step in understanding information flows in cells. The actomyosin cytoskeleton embeds controller and sensor elements which facilitate a flow of mechanical signals.\cite{debellyInterplayMechanicsSignalling2022} Upstream kinases and phosphatases regulate the activity of non-muscle myosin, and mechanosensitive proteins trigger downstream signal cascades when stretched. Such components likely endow cells with the controllability and observability needed to establish full-state feedback control.\cite{bechhoeferControlTheoryPhysicists2021b} The actomyosin cytoskeleton can be viewed not only as a generator of force, but also as a conduit of mechanical information from upstream regulatory pathways to downstream mechanosensitive cascades. This view is emerging in similar active and biological systems.\cite{najmaCompetingInstabilitiesReveal2022,chennakesavaluAdaptiveNonequilibriumDesign2024,davisActiveMatterControl2024} Control over contractility and signal throughput in reconstituted actomyosin active gels will therefore not only shed light on cellular processes, but could also be used in designs for micron-scale robots or synthetic cells.\cite{jia3DPrintedProteinbased2022,baldaufReconstitutedBranchedActin2022,reverte-lopezSelforganizedSpatialTargeting2024}

\section*{Conclusion}
In this study, we have leveraged concepts from control theory to study and characterize the role of variable inputs in contractile actomyosin systems. We demonstrate the presence of inherent nonlinearities for step, and quasistatic impulse inputs. In addition to providing insights into living systems, purified actomyosin active gels could be leveraged as actuators in bio-inspired microrobots or synthetic cells.\cite{jia3DPrintedProteinbased2022,baldaufReconstitutedBranchedActin2022} Indeed, the actomyosin cytoskeleton in living cells has many advantages compared to most actuators. Our results help to shift the framing of inquiry towards thinking of actomyosin's mechanical role in cellular signal, and information processing. Future work can begin to propose inquiries about the nature of feedback in these gels. In the way that a proportional-integral-derivative controller regulates the throttle of an automobile to maintain a desired velocity, we ask if there might be ways to regulate the chemical free energy available to our active gels to achieve certain physical responses? 

% \nocite{*} % This will include all entries from your .bib file
\section*{Conflicts of Interest}
The authors report no conflicts of interest.

\section*{Data Availability}
Data for this article are available at the Texas Data Repository at [URL to be added upon acceptance].

\section*{Acknowledgments}
This research was supported in part by the National Science Foundation through the Center for Dynamics and Control of Materials: an NSF MRSEC under Cooperative Agreement No. DMR-1720595 and DMR-2308817. This research was supported in part by grant NSF PHY-2309135 and the Gordon and Betty Moore Foundation Grant No. 2919.02 to the Kavli Institute for Theoretical Physics (KITP). We thank Eric Ansyln for providing insight, and materials for the use of caged ATP in this study.

%Here you should list the contents of your Supplementary Materials -- below is an example. 
%You should include a list of Supplementary figures, Tables, and any references that appear only in the SM. 
%Note that the reference numbering continues from the main text to the SM.
% In the example below, Refs. 4-10 were cited only in the SM.     
\section*{Supplementary materials}
Methods\cite{currentmanuscript_SI}\\
% References \textit{(\cite{reesBasicEngineeringPlasticity2012,bisswangerChapterReactionOrder,hubleyDiffusionCoefficientsAtp1995,koelschDiffusionMRHyperpolarized2013,siritanaratkulTransferPhotosyntheticNADP2017,dantzig18StudiesMolecular1998,ellis-daviesCagedCompoundsPhotorelease2007,goldmanInitiationActiveContraction1984,thirlwellKineticsRelaxationRigor1994,walkerPhotolabile2nitrophenylEthyl1988,kubitschkeActinMicrotubuleNetworks2017,romet-lemonneMechanotransductionIndividualActin2013,zemelMechanicalConsequencesCellular2011,liverpoolMechanicalResponseActive2009,nanningaKineticConstantsInteraction1960,kovacsFunctionalDivergenceHuman2003,sheetzATPdependentMovementMyosin1984a,philipsHowMuchEnergy})}

\section*{Author contributions}
J.C. and F.C. contributed equally to this work. J.A. conceptualized the work; J.C., F.C., and A.D. conducted the experiments. J.C., F.C., A.M., and J.A. contributed to modeling results. J.C. and J.A. led writing efforts.

\bibliographystyle{unsrt}
\bibliography{main}

\clearpage

\end{document}

\let\oldthebibliography=\thebibliography
\let\oldendthebibliography=\endthebibliography
\renewenvironment{thebibliography}[1]{
    \oldthebibliography{#1}
    \setcounter{enumiv}{57}                        % Change this number as required
}{\oldendthebibliography}
\title{Supplementary Material}% Force line breaks with \\

\maketitle

%\tableofcontents

\section*{Methods}

\subsection*{Slide preparation}
Glass slides are cleaned by sonication in deionized water for five minutes, followed by immersion in a piranha solution composed of five parts deionized water, one part $>30 \%$ ammonium hydroxide, and one part $>25\%$ hydrogen peroxide at $80 ^{\circ}\text{C}$ for thirty minutes. After the piranha solution, the slides are sonicated in deionized water for five minutes, and then dried and stored in an isopropanol solution. When a fresh slide set is desired, the slide and coverslip are taken out of isopropanol and dried. Parafilm is placed on the slide, and strips approximately 2.5 mm wide are cut from the parafilm using a sterile scalpel. The coverslip is cut into a strip $\sim$ 2.5 mm wide, and placed on top of the parafilm. The glass slide, parafilm, and coverslip are then placed on a hot plate at $120^{\circ}$C such that the parafilm melts, bonding the slide and coverslip together. The coverslip is gently pressed to ensure the parafilm bonds to the slide.

\subsection*{Sample preparation - NADH}
Actin, fascin, myosin are centrifuged at 100,000 g to precipitate aggregated, non-functional proteins. The concentration of each protein is determined with a Nanodrop 2000 Spectrophotometer. Fascin and myosin are added to a sample buffer with a concentration of 50 mM KCl, 2mM $\mathrm{MgCl_2}$, 20 mM imidazole, 0.1 mM ATP, 1 mM DL-dithiothreitol, 2 mM protocatechuic acid, \qty{0.1}{\micro M} protocatechuase 3,4-dioxygenase, 1 mM nicotinamide adenine dinucleotide, 2 mM phosphoenol pyruvate, and 15 units/mL or more of a pyruvate kinase (PK) and lactate dehydrogenase (LDH) mix from Sigma Aldrich, product number P0294. Actin is prepared in solution at $[\mathrm{actin}]$ = \qty{12}{\micro M}. The molar ratio of myosin to actin is varied between $\frac{1}{50}$ and $\frac{1}{200}$. The molar ratio of fascin to actin is varied between $\frac{1}{10}$ and $\frac{1}{20}$.

\subsection*{NADH calibration and ATP free energy}
The PK-LDH reactions stoichiometrically couple ATP consumption to NADH oxidation. In order to correlate a loss of NADH fluorescence with ATPase activity, calibration is required. To this end, we perform a dilution series of samples containing known concentrations of NADH and image with the same excitation intensities and exposure times as in experiment. We plot intensity versus concentration and observe a largely linear relationship. We fit this relation to a line to recover a conversion from measured fluorescence intensity to inferred NADH concentration in experiment.

\subsection*{Sample preparation - pulsed}
Actin was purified from rabbit psoas skeletal muscle from Pel-Freeze using a GE Superdex 200 Increase HiScale 16/40 column and stored at $\SI{-80}{\degreeCelsius}$ in G-Buffer ($\qty{2}{\milli M}$  tris-hydrochloride pH 8.0, $\qty{0.2}{\milli M}$ disodium adenosine triphosphate (ATP), $\qty{0.2}{\milli M}$ calcium chloride, and $\qty{0.2}{\milli M}$ dithiothreitol). All protein stocks were clarified of aggregated proteins at 100 000 gg for five minutes upon thawing and used within seven days. The G-actin concentration in the supernatant was determined by measuring the solution absorbance at $\qty{290}{\nano\meter}$ with a NanodDrop 2000 (Thermo Scientific, Wilmington, DE, USA) and using an extinction coefficient of $\qty{26600}{M^{-1}\centi\meter^{-1}}$. Actin is dialyzed in monomeric buffer containing \qty{0.001}{mM} ATP concentration to prevent large pre-contraction events due to residual ATP present in the standard actin monomeric storage buffer.

In addition to the standard contractile assay reagents (else ATP), pulsed excitation experiments utilize NPE-caged ATP (CATP), and phalloidin. Assays are formulated with $\left[\mathrm{CATP}\right] = \qty{0.1}{m M}$, and $[\mathrm{Phalloidin}] = \qty{12}{\micro M}$. Note that mixed assay are added to evaporated phalloidin as the final mixing step before loading into the chamber. Phalloidin is stored in methanol which will destroy the actin network if not sufficiently evaporated using a clean nitrogen line. Samples are labeled at a labeling ratio of $R=0.05$.

\subsection*{Imaging protocol}
The samples are imaged in a Zeiss Axio Observer 5 epifluoresence microscope with a Teledyne Photometrics Prime BSI Scientific CMOS camera and Lumen Dynamics X-cite Xylis LED fluorescence microscope. We image using mPlum (actin) and DAPI (caged-ATP release) filters to provide two optical channels. The DAPI filter excites NADH, and the PLUM filter excites the AlexaFluor dye bonded to actin filaments. For NADH experiments, we illuminate at 15\% (DAPI) intensity and 60\% (mPlum) intensity with an exposure time of \qty{5}{\milli\second} and \qty{50}{\milli\second}, respectively. Images are acquired every $20s$ for both channels. Each experiment is imaged for a total of 30 minutes. For CATP experiments, we illuminate at 100\% (DAPI) intensity and 60\% (mPlum) intensity with an exposure time of 50-\qty{500}{\milli\second} and \qty{2.5}{\milli\second}, respectively. mPlum images are acquired every $20s$. Pulses of UV light are applied after 45 mPlum image captures, for 15 pulses. Each experiment is imaged for a total of 225 minutes. For high throughput parallel experiments, a custom Python-based control system is utilized based on the OAD platform from Zeiss.

% \subsection*{Imaging and analysis}
% Imaging is performed on a Zeiss Axio-observer inverted epifluorescence microscope with Teledyne Photometrics Prime BSI Scientific CMOS camera and Lumen Dynamics X-cite Xylis LED fluorescence microscope light source. We image using mPlum (actin) and DAPI (caged-ATP release) filters with a EC Plan-NEOFLUAR 2.5x NA 0.085 objective (Zeiss). mPlum intensity is set to $60$\% with $2.5ms$ exposure. mPlum images are acquired every $20s$. DAPI intensity is set to $100$\% with variable exposure from $500ms$ to $10s$. For high throughput experiments, a custom Python based control system is utilized based on the OAD platform from Zeiss. Analysis is performed using Python with standard libraries.

\subsection*{Image processing - NADH}
We process the images from each experiment with a custom workflow based on Fiji/ImageJ and Python to extract actin strain and NADH consumption (Fig.(\ref{fig:CodeFlow})). First, we record two series of fluorescence images: labeled actin (PLUM channel; Fig.(\ref{fig:CodeFlow}a)) and NADH (DAPI channel; Fig.(\ref{fig:CodeFlow}j); cf. Imaging Protocol). We apply Otsu binarization alongside a flood fill algorithm (Fig.(\ref{fig:CodeFlow}b,e)). This step is to determine the perimeter of the gel. In the early stages of each experiment, it is sometimes difficult to achieve high enough contrast to determine this perimeter (Fig.(\ref{fig:CodeFlow}b)). In these cases, we resort to manual intervention by producing binary images using the "polygon selections" tool in Image J (Fig.(\ref{fig:CodeFlow}c,d)). If the image binarizes without error, the image is processed without manual infilling (Fig.(\ref{fig:CodeFlow}e)).

We next calculate the strain from the binarized images. We sum the imaged area of all thresholded pixels to determine gel imaged area (Fig.(\ref{fig:CodeFlow}f)). Next, we determine the Hencky strain using the equation for $\varepsilon$ found in “Hencky strain” (Fig.(\ref{fig:CodeFlow}g)). Integrating yields strain trajectories $\varepsilon (t)$ (Fig.(\ref{fig:CodeFlow}h)). 

The DAPI channel tracks NADH consumption (Fig. \ref{fig:CodeFlow}j). We use the calibration curve (Fig.(\ref{fig:CodeFlow}k)); see NADH Calibration and ATP Free Energy) to convert pixel fluorescence intensities from each image to NADH molar concentration (Fig.(\ref{fig:CodeFlow}l)).

To calculate total NADH consumption, we consider pixels within a region of interest determined by two factors. First, for each pair of DAPI-PLUM images, we apply a mask of the binarized PLUM image (corresponding to the imaged area of the actomyosin gel) to the corresponding DAPI image. This step allows us to exclude extraneous DAPI signals from the vacuum grease, which we have found exhibits auto-fluorescence under UV light. Second, in certain cases, we apply a manually determined mask to eliminate additional fluorescence signals originating from outside the sample. We then compute the total change in NADH concentration (Fig.(\ref{fig:CodeFlow}m)).

The change in NADH observed is primarily due to ATP hydrolysis by myosin motors, but there is also a non-negligible amount of photobleaching. To determine the photobleaching constant, we image an NADH-only sample under the same conditions as experiment.

We include a diagram of the image processing workflow in Fig.(\ref{fig:CodeFlow}).

\begin{figure}[!ht]
    \centering
    \includegraphics[width=0.95\textwidth]{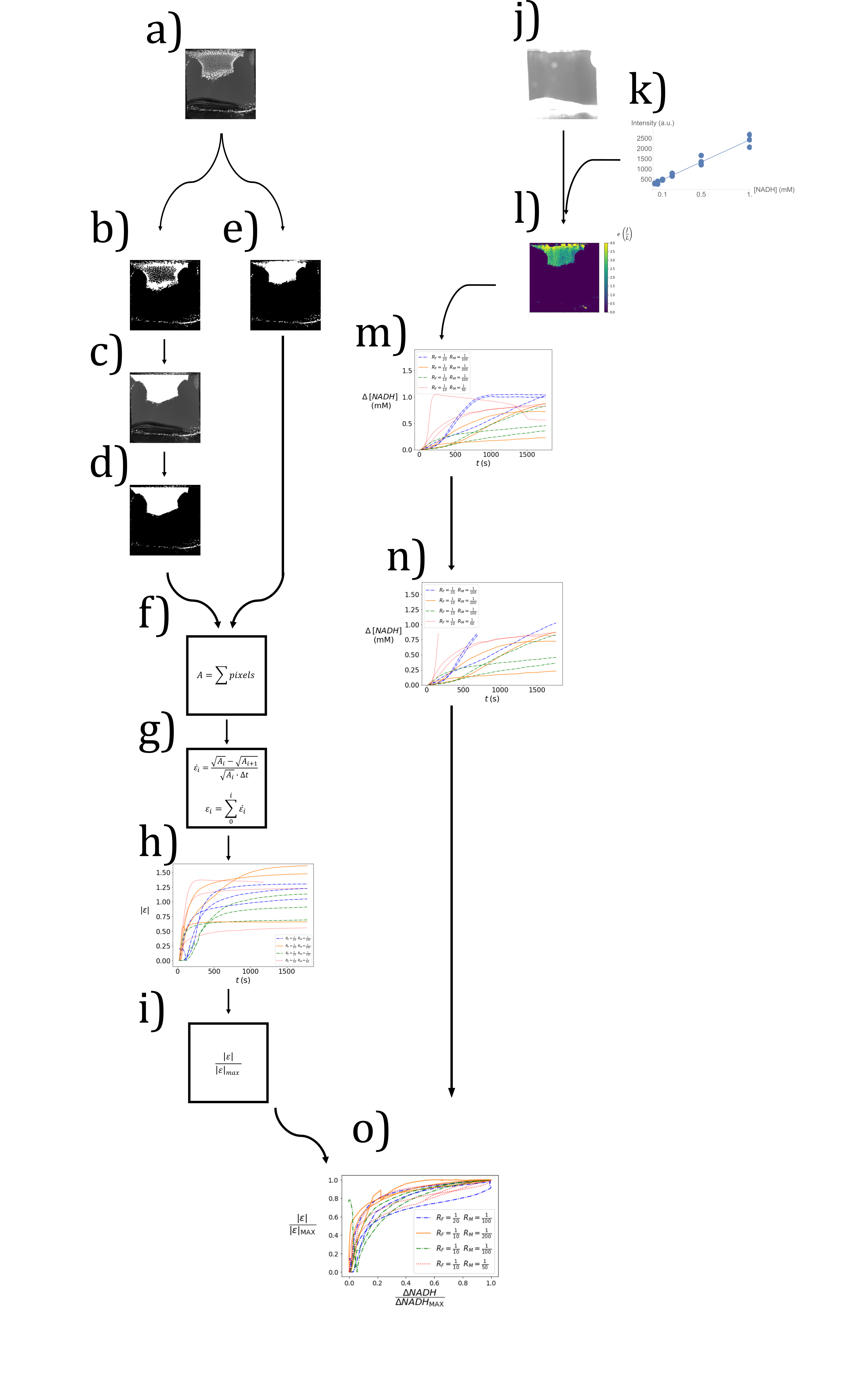}
\end{figure}

\clearpage % Forces the rest of the figure's content to the next page

\captionof{figure}{Processing strain and energy requires a set of complex steps a) The initial image of the actomyosin gel is subdivided into its frames such that each frame is processed individually. b) An otsu binarization algorithm is applied, followed by a flood-fill algorithm. Each image is examined individually to check if the binarized pixels span the whole area occupied by the actin mesh. c) If an image is found to be unsatisfactory, the original frame is manually filled in around the edges. d) The same otsu binarization algorithm and flood-fill algorithm is applied. e) Most images do not require the processing detailed in step b. These images are accepted automatically. f) To obtain the area of the mesh for an individual frame, the total white pixels in the image are counted. g) The square of these pixels is taken to obtain a characteristic length. Sequential lengths are then subtracted from each other and divided by the largest characteristic length between them and the timestep to obtain a change in Hencky Strain for that timestep. h) The absolute value of these strain measurements are computed. i) From the absolute value of these strain measurements, the maximum value of strain reported is selected for each dataset of strain, and used to normalize the each dataset of strain. j) The original NADH channel, which measures energy consumption, is divided into its constituent frames. k) A calibration curve mapping known NADH concentrations to the reported intensity of fluorescence is used to calculate total NADH molar consumption within the gel. In this experiment, NADH consumed directly corresponds to ATP rephosphorylation on a 1-to-1 ratio, which makes it a good indirect parameter of free energy consumption. l) The total NADH consumption is mapped onto the original frame using the calibration curve and the binarized location of the actin mesh network. m) The total NADH consumption per volume is measured as the average of the energy consumption per volume across the entire image. This average is then plotted against time. n) Cases where the reported NADH consumption saturates before the end of the experiment are truncated below a value of 1.0 M, to ensure that only valid measurements are reported for the actomyosin gel. o) Normalized strain is plotted against normalized NADH consumption.}\label{fig:CodeFlow}

\subsection*{Image processing - pulsed}
From each dataset, the final image of each pulse capture window (image 45) is selected for inter-pulse analysis. These inter-pulse subset representations of the experiment are converted into binarized images and then convert to gel area. Next, we determine the Hencky strain using the equation for $\varepsilon$ found in “Hencky strain” below. Analysis is performed using Python with standard libraries.

\subsection*{Hencky strain}
The Hencky strain $\varepsilon$ is given by \cite{reesBasicEngineeringPlasticity2012}

\begin{equation*}
    \varepsilon = \int \frac{dl}{l}
\end{equation*}

where the integral is performed over a sequence of incremental changes in line elements $dl$ with respect to their lengths $l$. This quantity is preferred over the engineering strain for large deformations, typically above 1\% strain, since for deformations larger than 1\% Hencky strain is path-independent.\cite{reesBasicEngineeringPlasticity2012} By taking length $l = \sqrt{A}$ as the square root of the imaged area of the gel, and discretizing $dl \approx \Delta l$, we have:

\begin{equation*}
\varepsilon = \int \frac{dl}{l} \approx \sum \frac{\Delta l}{l_i} = \sum
\frac{l_{i} - l_{i+1}}{l_i} = \sum
\frac{\sqrt{A_i} - \sqrt{A_{i+1}}}{\sqrt{A_i}}
\end{equation*}

\section*{Reaction-diffusion simulation}

To model the dynamics of free-energy consumption during contraction events, we perform a numerical simulation of the reaction-diffusion dynamics of the system. We begin with the enzymatic reactions that couple free-energy consumption to fluorescent NADH oxidation via myosin (Myo), pyruvate kinase (PK), and lactate dehydrogenase (LDH):

\begin{equation*}
\begin{aligned}
\ce{ATP &\xrightarrow{\rm Myo} ADP + P_i} \\
\ce{ADP + PEP &\xrightarrow{\rm PK} ATP + Pyr} \\
\ce{Pyr + NADH &\xrightarrow{\rm LDH} NAD^+ + Lac}
\end{aligned}
\end{equation*}

We thus have the following reaction equations for five reagents: adenosine triphosphate (ATP), adenosine triphosphate (ADP), phosphoenolpyruvic acid (PEP), pyruvate (Pyr), and nicotinamide adenine dinucleotide (NADH).

\begin{equation*}
\begin{aligned}
\dot c_{\rm ATP} &= -k_{\rm Myo} c_{\rm ATP} + k_{\rm PK}c_{\rm ADP}c_{\rm PEP} \\
\dot c_{\rm ADP} &= +k_{\rm Myo} c_{\rm ATP} - k_{\rm PK} c_{\rm ADP}c_{\rm PEP} \\
\dot c_{\rm PEP} &= -k_{\rm PK} c_{\rm ADP}c_{\rm PEP} \\
\dot c_{\rm Pyr} &= +k_{\rm PK} c_{\rm ADP} c_{\rm PEP} - k_{\rm LDH} c_{\rm Pyr} c_{\rm NADH} \\
\dot c_{\rm NADH} &= -k_{\rm LDH} c_{\rm Pyr} c_{\rm NADH} - k_b c_{\rm NADH}
\end{aligned}
\end{equation*}

where $c$ represents a reagent’s molar concentration, $\dot c$ represents its time derivative, $k$ represents an enzyme’s rate constant, and $k_b = \qty{163}{s^{-1}}$ represents the NADH photobleaching rate constant. We treat the myosin hydrolysis reaction as a pseudo-first-order reaction (neglecting water), while the PK and LDH reactions are considered as second-order. We assume the initial substrate-enzyme binding kinetics are much faster than the catalytic rates, allowing us to work in the enzymatic steady state.\cite{bisswangerChapterReactionOrder} Furthermore, we assume the dynamics of inorganic phosphate $\ce{P_i}$, oxidized NADH ($\ce{NAD^+}$), and lactate (Lac) waste products to be negligible.

Having modeled the reactions of the system, we next model the spatial diffusion of the reagents via the diffusion equation:

\begin{equation*}
    \dot c_m = D_m \nabla^2 c_m
\end{equation*}

where the index $m$ represents one of the five reagents, $D$ the diffusion constant, and $\nabla^2$ the Laplacian. We explicitly model diffusive dynamics in two-dimensions.

Finally, we assume that myosin ATPase occurs only within the actomyosin gel, which contracts over time; as the gel contracts, the density of myosin motors increases to conserve total myosin number. We model the gel as a thin, right square prism $\mathcal{S}$, anchored at the top-left corner of the system, whose base lengths $\ell$ (and height $h$) decrease as contraction proceeds. The ATPase rate $k_{\rm gel}(x,y,t)$ of the gel is zero outside of $\mathcal{S}$, while inside $\mathcal{S}$ it scales with $\lambda(t) = \ell(t)/\ell(0)$ to reflect increasing myosin density:

\begin{equation*}
    k_{\rm gel}(x,y,t) = \left\{
    \begin{matrix}
    k_{\rm Myo} \lambda(t)^{-3} & (x,y) \in \mathcal{S} \\
    0 & \text{otherwise} 
    \end{matrix}
    \right.
\end{equation*}

Furthermore, we assume that PK and LDH concentrations remain  constant in space. Combining the above equations by adding terms yields the following set of reaction-diffusion equations for the spatiotemporally resolved $c_m$ and $k_{\rm gel}$:

\begin{equation*}
\begin{aligned}
\dot c_{\rm ATP} &= D_{\rm ATP} \nabla^2 c_{\rm ATP} -k_{\rm gel} c_{\rm ATP} + k_{\rm PK}c_{\rm ADP}c_{\rm PEP} \\
\dot c_{\rm ADP} &= D_{\rm ADP} \nabla^2 c_{\rm ADP} +k_{\rm gel} c_{\rm ATP} - k_{\rm PK} c_{\rm ADP}c_{\rm PEP} \\
\dot c_{\rm PEP} &= D_{\rm PEP} \nabla^2 c_{\rm PEP} -k_{\rm PK} c_{\rm ADP}c_{\rm PEP} \\
\dot c_{\rm Pyr} &= D_{\rm Pyr} \nabla^2 c_{\rm Pyr} +k_{\rm PK} c_{\rm ADP} c_{\rm PEP} - k_{\rm LDH} c_{\rm Pyr} c_{\rm NADH} \\
\dot c_{\rm NADH} &= D_{\rm NADH} \nabla^2 c_{\rm NADH} -k_{\rm LDH} c_{\rm Pyr} c_{\rm NADH} - k_b c_{\rm NADH}
\end{aligned}
\end{equation*}

Our reaction-diffusion equations neglect advection of solvent during contraction. Furthermore, we restrict diffusion to two dimensions, and neglect diffusion in the $z$-direction. Therefore, the diffusion constants in our model are only approximations to the actual dynamics of the reagents in our system. In order to semi-empirically account for differences between simulation and experiment, we express all five diffusion constants in our model in terms of an effective, two-dimensional diffusion constant $D_{eff}$ and allow this one parameter to vary when fitting model to experiments (see below). We set $D_{\rm ATP} = D_{\rm ADP} = D_{eff}$ (given that the diffusion constants of ATP and ADP are comparable),\cite{hubleyDiffusionCoefficientsATP1995} $D_{\rm PEP} = D_{\rm Pyr} = 3.16 D_{eff}$,\cite{koelschDiffusionMRHyperpolarized2013} and $D_{\rm NADH} = 1.89 D_{eff}$. \cite{siritanaratkulTransferPhotosyntheticNADP2017}

In order to numerically simulate our model, we spatially resolve the concentration fields $c_m(x,y,t)$ at time $t$ by 75-by-75 matrices. The system size is $(\qty{1570}{\micro m})^2$. We initiate the $c_m$ with our experimental initial conditions.

We integrate the dynamic equations forward in discrete time using Euler’s method. Given $\dot c = f(c)$ and two times $t_i$ and $t_{i+1} = t_i+\Delta t$, we have $c(t_{i+1}) = c(t_i) + \Delta t f(c(t_i))$. We set the time step $\Delta t$ to $\qty{0.1}{s}$. We approximate the two-dimensional Laplacian $\nabla^2 = \partial_x^2 + \partial_y^2$ with the five-point stencil method. The final result, after simulating for 30 min, is the simulated concentration field $c_{\rm NADH}(x,y,t)$ of NADH, which we compare to experiment.

% \begin{figure}[!ht]
%     \centering
%     \includegraphics[width=0.75\textwidth]{figures_si/SimulationResults_02.png}
%     \caption{Simulation principles and results. a) A collection of the simulation 75x75 grid showing NADH concentration at nine different points during its evolution. The red lines in the bottom right corner represent the horizontal, vertical, and diagonal traces averaged together to capture the NADH kymograph in simulation. b) A kymograph of an experiment at 200 $s$ time intervals. c) A sample of the simulation (lines) compared to the observed NADH values according to best-fit values determined by the Nelder-Mead algorithm. d) The reported myosin activity values from the Nelder-Mead algorithm for all twelve datasets. They are in order of ascending myosin concentration: $R_M = \frac{1}{200}$: $k_{my} = 1.4\cdot 10^{-3} \pm 6.1 \cdot 10^{-5}$ $\frac{\mathrm{mM}}{\mathrm{min}}$, $R_M = \frac{1}{100}$ (pooled): $k_{my} = 4.4\cdot 10^{-3} \pm 3.2 \cdot 10^{-3}$ $\frac{\mathrm{mM}}{\mathrm{min}}$, and $R_M = \frac{1}{50}$: $k_{my} = 0.011 \pm 6.6 \cdot 10^{-3}$ $\frac{\mathrm{mM}}{\mathrm{min}}$. e) The reported diffusion values from the Nelder-Mead algorithm for all twelve datasets. No statistical significance is observed between datasets for either the myosin activity values or diffusion values.}
%     \label{fig:NADHSimulation}
% \end{figure}

In order to extract the effective diffusion constant $D_{eff}$ and the myosin rate constant $k_{\rm Myo}$ from an experiment, we perform optimization according to the Nelder-Mead method, using our reaction-diffusion simulation as a nonlinear fit function. For each experiment, we initialize a simulation with initial guesses for $D_{eff,0}$ and $k_{\rm Myo,0}$, and run the simulation according to the reaction-diffusion equations for $1800 s$. In addition, we include experimentally determined $\lambda(t)$ from the PLUM channel. This yields a final NADH concentration field, which we compare to experiment.

\section*{Bleaching rate estimates}
While it is possible to calculate the rate of NADH bleaching from purely theoretical principles, a more direct approach can use the data found in experiment to estimate the bleaching rate with minimal assumptions. Assuming that bleaching follows exponential decay $[NADH] = [NADH]_0 \cdot e^{-k_{Bl}t}$, the rate of exponential decay is slow compared to the timescales over which the experiment is measured, and that the rate of pyruvate diffusion does not travel the width of the slide during the experiment's timeframe, we can calculate the bleaching rate of NADH using the furthest radial edge of our kymographs.

\section*{Control theory review} Control theory considers what inputs lead to desired outputs within a system rather than simply treating the system as an evolving physical process with a predetermined input. Control theory is a good candidate for parameterizing the transfer functions of reconstituted actomyosin gels. This "transfer function" is a function that takes an input into a system, and creates an output based on that input. This transfer function can be static in time or dynamically evolve. 

To give a simple mathematical example, consider the first order system given by,
\begin{equation}
    \dot{y}(t)=-y(t) + u(t),
\end{equation}
where $y(t)$ is the measurable output of a system, and $u(t)$ is the input. Applying a Laplace Transform to the system gives,
\begin{equation}
    s\,y(s)=-y(s)+u(s).
\end{equation}
In Laplace space, control theory permits the following relationship,
\begin{equation}
    T(s) = \frac{\mathrm{Output}}{\mathrm{Input}} = \frac{y(s)}{u(s)} = \frac{1}{1+s},
\end{equation}
hence defining the transfer function for the system, $T(s)$.

\section*{Impulse response results}

First, we briefly define a transfer function. For a system that takes one input, $u(t)$, and generates an output, $y(t)$, based on that input. The transfer function of this system, $T(s)$, can be defined in Laplace space via $T(s)=\mathcal{L}(y(t))/\mathcal{L}(u(t))=y(s)/u(s)$, where $\mathcal{L}$ is the Laplace transform, and $s$ is frequency. For a first order model system, the high frequency response of the system scales as $s^{-\xi}$ where $\xi=1$ (SI). In general, the ability to define a transfer function for a system requires a linear time-invariant (LTI) system. For this reason, we argue that the transfer function approach doesn't work well for step input data. In the step input case, the system has age-dependence, is not time-translationally invariant, and exhibits many physical effects including nonlinearity. The advantage of pulsatile, quasistatic driving is that we can probe the nonlinear response of the system directly to small perturbation.

With this result, we are beginning to elucidate how the linear response of a nonlinear system changes with increasing input. If we appeal to the field of rheology, for conventional, externally loaded materials, this would be accomplished by measuring the so-called tangent compliance $K^{-1} = \delta\varepsilon/\delta\sigma$, that is, how the system responds to a strain profile $\sigma(t) = \sigma_0 + \sigma_1 \sin{\omega\,t}$, with $\sigma_1 \ll \sigma_0$. Measuring the resulting compliance $K^{-1}$ around different operating points $\sigma_0$ produces a curve $K^{-1}(\sigma_0)$ that accurately describes the nonlinear, pre-stress response of the system. In analogy, here we measure $\delta\varepsilon/\delta g$. Unlike with rheology, we cannot perform oscillatory rheology on myosin motors, because actomyosin gels only contract and do not expand. However, our protocol of delivering a series of pulses allows us to measure the “relaxation” of the system to a small input pulse of energy $\delta g$.

\subsection*{Description of pulsed excitation model}

% In order to contextualize the results of the current work, we begin with an explanation of our system's control diagram (Fig.(\ref{methodscontroldiagram}a)). A standard control diagram gives an input $u$, and output $y$ that are related to each other through the system transfer function $G$ such that,

% \begin{equation*}
% y = G\,u \implies G=\frac{y}{u}\,.
% \end{equation*}

% \begin{figure}[!ht]
% 	\centerline{\includegraphics[width=0.75\linewidth]{figures_si/transfer_function_global_v3_crop.png}}
% 	\caption{{(a) Simplified control diagram showing input, $u$, and output $y$, as related by the system transfer function A. (b) Proposed control theory diagram for our active gel system.}}\label{methodscontroldiagram}
% \end{figure}

Our initial input, $u$, for our system is energy from UV photons. We incorporate the well-studied reagent, NPE-caged ATP (Adenosine 5'-Triphosphate, P3-(1-(2-Nitrophenyl)Ethyl) Ester, Disodium Salt) purchased from Thermo Fisher Scientific, into our assay which prevents myosin II ATPase activity.\cite{dantzig18StudiesMolecular1998,ellis-daviesCagedCompoundsPhotorelease2007,goldmanInitiationActiveContraction1984,thirlwellKineticsRelaxationRigor1994,walkerPhotolabile2nitrophenylEthyl1988} The NPE cages can be destroyed via UV radiation in a dose-dependent manner wherein the does is set by the experimenter as a true input to the system. An input pulse is specified by a microscope set intensity value given as a dimensionless percentage. These intensity values correspond to a given Power $P$ as measured by a power meter. The other user input is exposure time $\tau_e$. Using, $P\times\tau_e [=]\,\mathrm{Joules}$, we can know the exact amount of photon energy delivered to the sample. 

The concentration of photoreleased NPE-caged ATP during photolysis is given by
\begin{equation}\label{eq_photolysis}
[ATP]_R = \left(I_0\,q_e\,\lambda\,[CATP]\,\epsilon\right)/\left(A\,h\,c\,N_A\right)
\end{equation}

where $I_0$ is the photolysis pulse energy, $A$ and $h$ are the area and depth of the sample chamber, $q_e=0.63$ is the quantum yield for photolysis,\cite{walkerPhotolabile2nitrophenylEthyl1988} $[CATP]$ is the concentration of NPE-caged ATP molecules, $\epsilon=660M^{-1}cm^{-1}$ is the molar absorption coefficient at $\lambda$,\cite{walkerPhotolabile2nitrophenylEthyl1988} $h$ is Planck's constant, $c$ is the velocity of light, and $N_A$ is Avogadro's number.\cite{dantzig18StudiesMolecular1998}

Our choice of output for our system is strain, $\varepsilon$. This choice of response variable is biologically relevant, as cells undergo measurable strains as they interact with their environment.\cite{kubitschkeActinMicrotubuleNetworks2017,romet-lemonneMechanotransductionIndividualActin2013,zemelMechanicalConsequencesCellular2011} It is typical to measure strain in response to an external stress, such as in rheology where the strain response of a material is measured in response to an applied shear. The rheological approach to studying reconstituted actin gels has been heavily utilized over the past few decades.\cite{liverpoolMechanicalResponseActive2009} Strain is a natural choice of response as it gives a measure on the change in bulk properties due to changes in the morphology of a contracting gel. Further, it is dependent on the transduction of chemical free energy to contractile work by myosin II molecular motors, more like an internal stress.

In our system, photon energy is converted into a release of ATP with a quantum efficiency of $q_e = 0.63$, as given by the NPE-caged ATP manufacturer specification. A given pulse of UV light releases an amount of ATP, as given by Eq.(\ref{eq_photolysis}), and thus increases the amount of chemical free energy associated with this release of ATP.

Our model is predicated on Michaelis-Menten kinetics (MMK).\cite{bisswangerChapterReactionOrder} We incorporate ATP concentration dependent myosin ATPase activity, where - in MMK nomenclature - myosin is our enzyme, and ATP, our substrate. Myosin II binds ATP at the head group and performs a power stroke to process along actin filaments in the network through ATP hydrolysis.

To implement our MMK model, we use a combination of Python, and Mathematica. First we calculate ATP release profiles based on Eq.(\ref{eq_photolysis}) using Python. We also define the effective area of the gel, based on our area measurements of contractile gels. We declare the effective area ($A_{eff}$) as the area of the gel by which liberated ATP can be hydrolyzed by myosin II motors. This is the area of the gel $A_{gel}$, as observed before a given pulse, plus the diffusive area $A_{diff}$ as calculated from the known ATP diffusivity in buffer solution using $A_{diff} = D_{ATP}\times N_{img}\times t_{delay}$ where $D_{ATP}$ is the known ATP diffusion constant,\cite{hubleyDiffusionCoefficientsATP1995} $N_{img}$ is the number of images per pulse, and $t_{delay}$ is the delay time between subsequent image captures. The amount of available ATP available to the gel is the total amount of ATP released for a given photolysis step modulo the effective area of the gel described above. This way, we are not considering the ATP that is released in the surrounding fluid when performing our MMK calculations. These were calculated for all of the gels measured. The effective area modulated release profiles for each pulse type are then used as inputs into the MMK model.

The MMK model is defined with Mathematica's NDSolve. The system consists of equations for the concentrations of inorganic phosphate ($c_{Pi}$), adenosine diphosphate ($c_{ADP}$), adenosine triphosphate ($c_{ATP}$), the enzyme complex ($c_{M}$), and a logarithmic ratio $q$ related to the concentrations.

The Michaelis-Menten constant ($K_M$) and the initial myosin II concentration ($c_{M_0}$) are defined as:

\begin{equation*}
K_M = \frac{k_{M_{-1}} + k_{M_{+2}}}{k_{M_{p1}}},\,\,\,\,c_{M_0} = 10^{-7}.
\end{equation*}

The differential equations are:
\begin{align*}
	\frac{d c_{Pi}}{dt} &= \frac{k_{M_{+2}} \, c_{M_0} \, c_{ATP}(t)}{c_{ATP}(t) + K_M}, \\[0.5ex]
	\frac{d c_{ADP}}{dt} &= \frac{k_{M_{+2}} \, c_{M_0} \, c_{ATP}(t)}{c_{ATP}(t) + K_M}, \\[0.5ex]
	\frac{d c_{ATP}}{dt} &= -k_{M_{+1}} \, c_{M}(t) \, c_{ATP}(t) + \left( k_{M_{+2}} + k_{M_{-1}} \right) \frac{c_{M_0} \, c_{ATP}(t)}{c_{ATP}(t) + K_M}, \\[0.5ex]
	\frac{d c_{M}}{dt} &= -k_{M_{-1}} \, c_{M}(t) \, c_{ATP}(t) + \left( k_{M_{+2}} + k_{M_{-1}} \right) \frac{c_{M_0} \, c_{ATP}(t)}{c_{ATP}(t) + K_M}, \\[0.5ex]
	\frac{d q}{dt} &= \frac{d c_{ADP}/dt}{c_{ADP}(t)} + \frac{d c_{Pi}/dt}{c_{Pi}(t)} - \frac{d c_{ATP}/dt}{c_{ATP}(t)}.
\end{align*}
% The initial conditions for the system are:

% \eqs{c_{Pi}(0) = c_{Pi_0},\,\,c_{ADP}(0) = c_{ADP_0},\,\,c_{ATP}(0) = c_{ATP_0},\,\,c_{M}(0) = c_{M_0},\,\,q(0) = \log\left(\frac{c_{ADP_0} \, c_{Pi_0}}{c_{ATP_0}}\right).}

This system of equations models the consumption of ATP and the corresponding production of ADP and inorganic phosphate ($c_{Pi}$) during contraction, with myosin II dynamics captured by $c_{M}$. The expression for $q$ captures the overall changes in the logarithmic ratio of the concentrations of the involved species. The parameters $k_{M_{+1}}, k_{M_{+2}},$ and $k_{M_{-1}}$ are rate constants for the forward and reverse reactions, while $K_M$ represents the Michaelis constant.

For a given gel the model is initialized using a rate for $k_{+1}$ pulled from our simulation results where we measure an ATP consumption rate of $v = \qty{0.0000833}{mM/s}$, our nominal concentration of myosin II, and a measure for $K_M$.\cite{nanningaKineticConstantsInteraction1960} The other two rates are pulled from literature.\cite{kovacsFunctionalDivergenceHuman2003} The simulation begins by taking the first element from the array of that gels area modulated release of ATP due to UV photolysis. The model proceeds for $\qty{900}{\second}$ to agree with the interpulse time from experiments. During each iteration, the amount of ATP consumed by the gel is calculated and stored to an array. Any remaining ATP is also stored in an array. The model then proceeds with the next iteration, pulling the next element from the caged ATP release profile array. For iterations beyond the first, any remaining ATP from the previous window is added to the amount of released ATP. During each iteration the myosin II ATPase activity is scaled by a factor varying with the ATP concentration given by,\cite{sheetzATPdependentMovementMyosin1984a} 
\begin{align*}
	k_{M_{+1,i}} &= \left(\frac{c_{ATP_i}}{c_{CATP_{init}}}\right)k_{M_{+1,init}},\\[0.5ex]
k_{M_{-1,i}} &= \left(\frac{c_{ATP_i}}{c_{CATP_{init}}}\right)k_{M_{-1,init}},\\[0.5ex]
	k_{M_{+2,i}} &= \left(\frac{c_{ATP_i}}{c_{CATP_{init}}}\right)k_{M_{+2,init}}.
\end{align*}
For small pulse durations, the myosin II are less processive, and thus consume less ATP.\cite{sheetzATPdependentMovementMyosin1984a} The model proceeds for 15 pulse releases. The arrays containing the amount of ATP consumed per pulse, as well as the amount of ADP and Pi, are saved into .csv files.

Next, the ATP consumption arrays, as well as the amount of ADP and Pi are loaded into Python where the remainder of the analysis takes place. ATP consumption arrays are converted to free energy using a conversion factor that considers to relative concentrations of ADP and Pi that are present during each contraction step (pulse).\cite{philipsHowMuchEnergy} With the free energy consumption arrays prepared, we utilized Hencky strain to determine strain from our area measurements of the gels as described above. In order to calculate transfer functions, the free energy consumption arrays and the strain evolution arrays were z-transformed in Python.

\bibliographystyle{unsrt}
\bibliography{main}